%% file: 0_main.tex
\documentclass[runningheads, dvipsnames]{llncs}

\input{_packages}

\usepackage{graphicx}
\begin{document}
\title{Learning-to-Rank with Nested Feedback}
%
%
\author{Hitesh Sagtani\orcidID{0009-0003-6995-1912} \and
Olivier Jeunen\orcidID{0000-0001-6256-5814} \and
Aleksei Ustimenko\orcidID{0009-0006-4942-7779}}
\authorrunning{Sagtani et al.}
%
\institute{ShareChat \\
\email{\{hiteshsagtani, jeunen, aleksei.ustimenko\}@sharechat.co}}

\maketitle              

\begin{abstract}
Many platforms on the web present ranked lists of content to users, typically optimized for engagement-, satisfaction- or retention- driven metrics.
Advances in the Learning-to-Rank (LTR) research literature have enabled rapid growth in this application area.
Several popular interfaces now include nested lists, where users can enter a 2\textsuperscript{nd}-level feed via any given 1\textsuperscript{st}-level item.
Naturally, this has implications for evaluation metrics, objective functions, and the ranking policies we wish to learn.
We propose a theoretically grounded method to incorporate 2\textsuperscript{nd}-level feedback into any 1\textsuperscript{st}-level ranking model.
Online experiments on a large-scale recommendation system confirm our theoretical findings.


\end{abstract}

\input{1_intro}
\input{3_concept}
\input{5_exps}
\input{7_discussion}

%
%
%
\bibliographystyle{splncs04}
\bibliography{references}

\end{document}

%% file: _packages.tex
\usepackage{graphicx} 
\graphicspath{{figures/}} 
\usepackage{multirow}
\usepackage{dsfont}
\usepackage{colortbl}
\usepackage[ruled,linesnumbered]{algorithm2e}

\usepackage{float}

\usepackage{caption}
\usepackage{subcaption}
\usepackage{booktabs}
\usepackage{amsmath}
\usepackage{amssymb}

\usepackage{wrapfig}

\usepackage{balance}

\usepackage{xcolor}

\usepackage{todonotes}
\makeatletter
\newcommand*\iftodonotes{\if@todonotes@disabled\expandafter\@secondoftwo\else\expandafter\@firstoftwo\fi}  
\makeatother

%% file: 1_intro.tex
\section{Introduction \& Related Work}

Rankings are at the heart of how users interact with content on the web.
This holds for applications and use-cases across web search and e-commerce recommendations to streaming and social media platforms.
Indeed, platforms use ranked list interfaces to serve content to users.
Machine learning models typically power these rankings, learned to optimize some metrics deemed relevant to the business or its users.
Due to this wealth of application areas, the Learning-to-Rank (LTR) literature has seen vast industry adoption~\cite{Karmaker2017,Pasumarthi2019,Haldar2020,Wang2021,Jagerman2022,JeunenFIRE2023}.

Nevertheless, real-world examples of ranking interfaces often deviate from the traditional setup where a single ranked list is shown.
Famous examples here include the \emph{gridwise} page layout that is now standardized in video streaming platforms~\cite{GomezUribe2016}, and the research literature has looked into more general interfaces as well~\cite{Oosterhuis2018,Xi2023}.
These works keep a broad focus, making few assumptions about the setting to provide effective methods with universal appeal.

\begin{figure}
    \centering
    \includegraphics[width=0.7\textwidth]{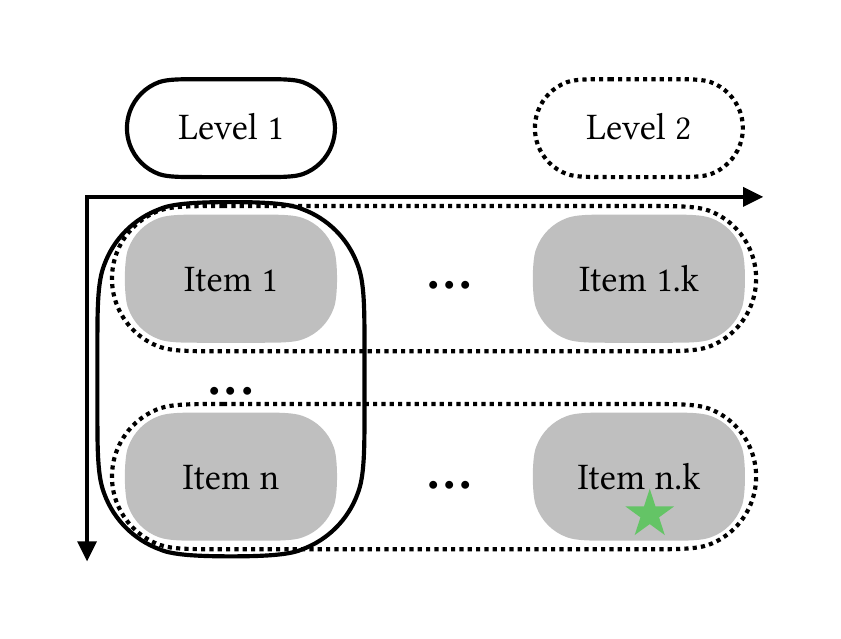}
    \vspace{-5ex}
    \caption{A nested-feed interface. When receiving feedback on item $n.k$ in the 2\textsuperscript{nd}-level feed (denoted \textcolor{Green}{$\star$}), the 1\textsuperscript{st}-level item $n$ should be attributed as well.
    }
    \vspace{-5ex}
    \label{fig:layout}
\end{figure}

Recently, \emph{nested} ranking lists have started to gain in popularity, particularly seeing adoption in short-video feeds on social media platforms such as Reddit, Instagram, and ShareChat~\cite{Jeunen_RecSys2023}.
Here, users are presented with a scrollable 1\textsuperscript{st}-level feed, where they can enter a full-screen 2\textsuperscript{nd}-level feed via any given 1\textsuperscript{st}-level item.
This differs from the grid layout discussed above, as only a single level is presented to the user at any given time.
Figure~\ref{fig:layout} visualizes such a layout of \emph{nested} feeds, which are the core topic of this paper.

A na\"ive modeling approach would instantiate both levels with independent ranking policies that optimize level-specific objectives.
This implicitly assumes that rewards across feeds are independent, which ignores the hierarchically nested structure.
Indeed, as shown in Figure~\ref{fig:layout}, positive feedback on 2\textsuperscript{nd}-level items should be (partially) attributed to the relevant 1\textsuperscript{st}-level item, as to allow the 1\textsuperscript{st}-level ranking policy to learn from this \emph{nested} feedback.

Existing methods in the literature that go beyond single lists either deal with complex settings where no assumptions are placed on the reward or examination dependencies in the interface~\cite{Oosterhuis2018}, or page-level re-ranking strategies when multiple independent lists are present~\cite{Xi2023}.
Even though nested ranking interfaces are prevalent in practical applications, they are currently under-explored in the research literature.
Our work formally introduces this problem setting.
Under the commonly assumed position-based model~\cite{chuklin2015click}, we theoretically derive the optimal objective for the 1\textsuperscript{st}-level ranking model when 2\textsuperscript{nd}-level nested feedback is available. 

\textbf{\textit{Related Work}}. Learning-to-Rank (LTR) is a classical problem in information retrieval that has received extensive research attention \cite{liu2009learning,li2011short}. It has found widespread adoption in various application areas, including web search \cite{qin2008learning}, question answering \cite{yang2016beyond}, e-commerce recommendations \cite{karmaker2017application}, streaming platforms \cite{Karmaker2017}, and recommendation systems \cite{duan2010empirical,karatzoglou2013learning,JeunenFIRE2023}.
LTR approaches can be categorized into different types, such as pointwise \cite{li2007mcrank}, pairwise \cite{leaman2013dnorm,burges2005learning,freund2003efficient}, and listwise \cite{xia2008listwise,cao2007learning,xu2007adarank,burges2010ranknet} methods.
Recent advancements in LTR include using online learning algorithms with non-linear models \cite{oosterhuis2018differentiable} and the proposal of stochastic bandit algorithms like BatchRank \cite{zoghi2017online}.
In the context of social media platforms, LTR algorithms are usually trained on implicit signals derived from user interactions, such as user clicks \cite{joachims2017accurately}, dwell time \cite{yi2014beyond}, and various other engagement signals \cite{lalmas2018tutorial}, that are then treated as indicators of relevance. However, these signals are subject to biases, such as positional bias, where the ranking position influences user clicks. To address this issue, many studies have explored methods to mitigate bias in ranking using counterfactual inference frameworks through empirical risk minimization \cite{joachims2017unbiased,ai2018unbiased}. Qingyao \textit{et al.} \cite{ai2021unbiased} evaluated and compared the performance of such state-of-the-art unbiased LTR algorithms.

%% file: 3_concept.tex
\section{Modeling Nested Ranking Signals}\label{sec:motivation}

In the rest of the work, we refer to the  1\textsuperscript{st}-level feed as L1 (Level 1), and the 2\textsuperscript{nd}-level feed as L2 (Level 2).
To calculate the final relevance label for a feed, we combine the following user signals using linear scalarization: \emph{likes}, \emph{shares}, \textit{fav}ourites, and \emph{video clicks}. The latter signal indicates that a user clicks on an L1 video to enter the L2 feed.
We observe the correlation among these signals to be significantly different for L1 and L2 feed for our platform, highlighting variations in user behaviours across feeds, resulting in different trade-offs between signals.

We consider some distribution of users $u\sim \mathcal{U}$ that interact with our platform.
Classical LTR deals with the problem of finding and evaluating a ranking for a list of items $A = \{a_{1}, \ldots, a_{n}\}$ with per-item relevance signals $y_{i} \forall i \in [1,n]$, e.g., a click or crowd-sourced label.
Typically, ranking quality is measured by Discounted Cumulative Gain (DCG):
\begin{equation}
    \mathrm{DCG} = \sum_{i=1}^{n} \frac{y_{i}}{\log_{2}(1+i)}.
\end{equation}
Given some model $f(u, a)$ we can sort items $a\in A$ in decreasing order to obtain a ranking.
LTR is then: finding a model $f(u, a)$ such that $\mathbb{E}_{u\sim \mathcal{U}}\mathrm{DCG}$ is maximized.
$\mathrm{DCG}$ objectives can be effectively optimized by state-of-the-art LTR approaches like LambdaRank \cite{burges2006learning}, StochasticRank \cite{ustimenko2020stochasticrank} or YetiRank \cite{gulin2011winning}.

Denote $R(u, a)$ as the online relevance signal that we observe for the ranked list.
Note that DCG adheres to the Position-Based Model (PBM) \cite{chuklin2015click}, when we assume that the probability of user viewing position $i$ is equal to $\frac{1}{\log_{2}(1+i)}$.
As a result, $y_i = \frac{R(u, a_{i})}{\mathbb{P}(u \text{ viewed position }i)}$ is an unbiased estimator of $\mathbb{E}\big(R(u, a_{i})|u, a, \text{viewed})$. 
Under these reasonable assumptions, $\mathrm{DCG}$ can be seen as an offline estimator of an online metric $Q$~\cite{jeunen2023normalised,jeunen2021offline}:
\begin{align}
    &Q = \mathbb{E}_{u\sim \mathcal{U}} \sum_{i=1}^{n} R(u, a_{i}) =\\
    &\mathbb{E}_{u\sim \mathcal{U}} \sum_{i=1}^{n} \mathbb{E}\big(R(u, a_{i})|u, a, \text{viewed})\mathbb{P}(u \text{ viewed position }i) = \mathbb{E}_{u\sim \mathcal{U}}\mathrm{DCG}.
\end{align}
In our case, when the user clicks on the item $a_{i}$, our system retrieves another \emph{nested} list of items $B_{i} = \{b_{i1}, \ldots, b_{im}\}$, ranks them, and presents them to the user. For an item $a \in A$ we denote as $R_{A}(u, a)$ a \emph{relevance} signal (e.g., an indicator of a positive interaction) that indicates user preference on the L1 feed, and we analogously define $R_{B}(u, b)$ for the L2 feed. The goal of the LTR model is to maximize the following online metric:
\begin{equation}
    Q = \mathbb{E}_{u\sim\mathcal{U}} \sum_{i=1}^{n} \big(R_{A}(u, a_i) + \sum_{j=1}^{m} R_{B}(u, b_{ij})\big).
\end{equation}
Suppose we ignore the contribution from $R_{B}$ and assume a classic setup with only $R_{A}$: items with lower L1 relevance but better L2 feeds are penalised and thus, a model trained only on L1 signals will have \emph{worse} online performance for metric $Q$.
This leads to a degradation of the recommendation system overall, while the ``quality'' of ranking on the L1 feed appears high. 

In our setup, we consider only how to train a ranker model to rank list $A$ so that $Q$ is maximized while assuming the ranking for $B$ is fixed. From the very definition of $Q$ it is seen that we should define $\widetilde{R}(u, a_i) = R_{A}(u, a_i) + \sum_{j=1}^{m} R_{B}(u, b_{ij})$ as the new relevance signal for the item $a_i$.
That is, the relevance signal for the L1 feed should account for the relevance signal on the L2 feed if we want to find a ranker that maximizes overall quality measured by $Q$.

Thus, to train a model we consider historical logs consisting of $(u, A, B, Y)$ where $Y = (y_{ij})_{i=1, j=0}^{n, m}$ denotes observed values for $y_{i0} \sim R_{A}(u, a_i)$ and $y_{ij} \sim R_{B}(u, b_{ij}), j>0$.
These logs are transformed into triplets $(u, A, \widetilde{Y})$, where we define $\widetilde{y}_i = \sum_{j=0}^{m} y_{ij}\sim \widetilde{R}(u, a_i)$. The dataset $\{(u, A, \widetilde{Y})\}$ now resembles a classic LTR dataset and, therefore, can serve as input to well-established LTR methods.

We also note that because the ranking on the L2 feed (i.e. the ranking of list $B$) is fixed, that means that we should not measure the contribution from the L2 feed using $\mathrm{DCG}$. This is because our observed reward in the logs $R_{B}(u, b_{ij})$ is already positionally biased. To see that, we can formally write:
\begin{align}
\mathbb{E}_{u\sim\mathcal{U}} &\sum_{j=1}^{m} R_{B}(u, b_{ij}) = \mathbb{E}_{u\sim\mathcal{U}}\frac{R_{B}(u, b_{ij})}{\mathbb{P}(u \text{ viewed position }j)}\mathbb{P}(u \text{ viewed position }j)
\end{align}
This means that we can get $\mathrm{DCG}$ if we consider debiased labels $\frac{R_{B}(u, b_{ij})}{\mathbb{P}(u \text{ viewed position }j)}$ but since the ranking of $B$ is fixed, those debiased labels have \emph{the same} denominator as the positional bias and, thus, it cancels out. Therefore, we should not discount observed relevance signals on $B$ and include them just as sums.

Our final feed-level relevance signals, denoted as $R_{A}(u, a_i)$ for the L1 feed and $R_{B}(u, b_{ij})$ for the L2 feed, are obtained by linearly combining individual user signals. These signals are fine-tuned internally to optimize user retention. However, since our objective is to evaluate the incorporation of user feedback from the L2 feed into the L1 feed, we do not delve into the specifics of how we tune the weights for each individual feed. In the subsequent section, we assign the following synthetic labels as the relevance signals for evaluation:
$S_1$ as $R_{A}(u, a_i)$, $S_2$ as $\widetilde{R}(u, a_i)$ with DCG of $R_{B}(u, b_{ij})$, and $S_3$ as $\widetilde{R}(u, a_i)$ with sum of $R_{B}(u, b_{ij})$.

%% file: 5_exps.tex
\section{Experimental Validation}
To the best of our knowledge, no datasets containing logged user feedback for nested feed interfaces are publicly available.
For this reason, we resort to a proprietary dataset obtained from ShareChat, a widely popular social media application with over 180 million monthly active users across 18 regional languages in India, to substantiate our hypothesis empirically.

We leverage a range of dynamic user and post attributes based on embeddings, user sign-up date, genre, tags, fatigue score \cite{sagtani2023quantifying} and language, along with several interaction-based attributes.
We find strong correlations between the relevance label with certain interaction features: the dot product between user and post embeddings is one such example.
We applied negative sampling to the dataset based on a union of all individual signals, i.e. \emph{likes}, \emph{shares}, \emph{favs} and \emph{clicks}. We collected a random sample of 100 million instances, where less than 5\% of the dataset had a positive signal, while the remaining instances were labeled as `0'. To ensure representative data for training, validation, and testing, we used stratified sampling, dividing the dataset in a 70:15:15 ratio, respectively.

\subsection{Offline Training and Experiment Results}\label{sec:offline_exp_results}
Gradient-Boosted Decision Tree (GBDT) methods such as LambdaMart have long remained the state-of-the-art approach \cite{qin2021neural} for tabular LTR datasets.
Recent empirical evaluations have shown YetiRank to outperform LambdaMart and other competing algorithms in most cases~\cite{lyzhin2023tricks}. To evaluate the synthetic labels, we use the YetiRank algorithm implemented in the Catboost~\cite{prokhorenkova2018catboost} library.
For hyperparameter tuning, we perform around 100 iterations of Bayesian optimization using the hyperopt~\cite{bergstra2013hyperopt} library for each synthetic label. To address the class imbalance problem in the dataset during training, we scale the loss of positive class by the ratio of negative to the positive examples using {\tt scale\_pos\_weight} parameter. We use an overfitting detector on the validation dataset and stop the training if there is no improvement after 50 iterations on \emph{DCG@10} and use the best-performing model on the validation set for each signal.

\begin{table}[!t]
\centering
\caption{Offline \% loss in DCG of predictor label compared to a model trained and ranked only on various ``true'' relevance signals.}
{
\begin{tabular}{ccccccccccc}
\toprule

\multirow{1}{*}{DCG} &~& \multirow{2}{*}{Label} &~& \multicolumn{7}{c}{\textbf{True Relevance Signal}} \\
\cline{5-11}
@k & &~&~& likes & shares & favs & clicks & $S_{1}$ & $S_{2}$ & $S_{3}$  \\
\midrule
\multirow{3}{*}{3} &~& $S_1$ &~& 27.8 & 26.8 & 23.7 & 17.3 & \textbf{0} & 17.3 & 18.1  \\
&~& $S_2$ &~& 24.1 & 25.2 & \textbf{20.5} & 14.8 & 10.2 & \textbf{0} & 15.1 \\
&~& $S_3$ &~& \textbf{20.7} & \textbf{24.7} & 20.8 & \textbf{13.4} & 12.4 & 8.4 & \textbf{0}  \\
\hline

\multirow{3}{*}{5} &~& $S_1$ &~& 25.9 & 25.8 & 20.3 & 15.6 & \textbf{0} & 16.5 & 17.7  \\
&~& $S_2$ &~& 22.9 & \textbf{24.1} & \textbf{17.5} & 14.4 & 10.1 & \textbf{0} & 15.2  \\
&~& $S_3$ &~& \textbf{19.2} & 24.3 & 18.2 & \textbf{12.6} & 10.5 & 8.1 & \textbf{0}  \\
\hline

\multirow{3}{*}{10} &~& $S_1$ &~& 24.5 & 24.2 & 19.6 & 15.1 & \textbf{0} & 16.1 & 17.3  \\
&~& $S_2$ &~& 20.8 & 23.6 & \textbf{16.7} & 11.6 & 9.3 & \textbf{0} & 14.4  \\
&~& $S_3$ &~& \textbf{17.8} & \textbf{23.2} & 17.4 & \textbf{10.8} & 10.4 & 7.7 & \textbf{0}  \\
\bottomrule
\end{tabular}
}
\vspace{-1em}
\label{tab:offline_results}
\end{table}

We evaluated the predictors ($S_1$, $S_2$, $S_3$) discussed in the previous section, considering \textit{likes}, \textit{shares}, \textit{favs}, \textit{clicks}, \textit{$S_1$}, \textit{$S_2$} and \textit{$S_3$} as the true relevance labels and evaluated on the \textit{DCG@k} metric. Table \ref{tab:offline_results} shows the \% loss in DCG metric values of predictor models compared to the model trained on the true relevance signal for ranking. For example, when using \emph{likes} as true relevance, we observe a 20.7\% decrease in \emph{DCG@3} metric for model trained on $S_3$ synthetic label, compared to the model trained on \emph{likes} signal.

From Table \ref{tab:offline_results}, we observe that models trained on $S_2$ and $S_3$ labels exhibit lower DCG loss compared to $S_1$ for all the individual user signals: \textit{likes}, \textit{shares}, \textit{favs} and \textit{clicks}, with $S_3$ having minimum loss across most of the signals.
This validates our hypothesis that incorporating L2 feedback leads to improved overall ranking, with a sum-aggregation of L2 feedback to be optimal over discounting, as outlined in Section \ref{sec:motivation}. 

We also treat each synthetic label as the \emph{true} relevance signal to compare them relatively. 
We note that the \% loss in DCG is much lower when using $S_3$ as predictor for other synthetic labels as the true relevance label ($S_1$, $S_2$), followed by $S_2$ and finally $S_1$ exhibiting the highest loss. Since, $S_3$ and $S_2$ predictors have both the L1 and L2 feedback information in their synthetic label, they are able to capture user behavior on both the feeds, compared to the $S_1$ predictor which has feedback for only L1 feed. In addition, the loss is much lower when using the $S_3$ predictor on $S_2$ as true relevance compared to the $S_2$ predictor with $S_3$ as true relevance, the $S_3$ label is better able to capture the information on L2 feed than $S_2$.

\subsection{Online A/B Experimentation}
Based on the offline results, we conducted an A/B experiment to evaluate various synthetic labels.
Table \ref{tab:online_ab} tabulates the online metrics for ranking based on $S_2$ and $S_3$ labels in \textit{variant-1} and \textit{variant-2} respectively, with $S_1$ being the control.

From Table \ref{tab:ab_feed_specific}, we observe that overall engagements (likes, shares, favs) have increased for both variants, indicating better posts on the L1 feed that lead to more engagements on the L2 feed. We also observe increased clicks on the L1 feed, indicating higher user convergence to the L2 feed compared to the model trained solely on the $S_1$ label. While there is a decrease in dwell time on the L1 feed, there is an increase for the L2 feed, suggesting that users spend more time overall in the L2 feed than in the L1 feed. The increase in clicks on the L1 feed also supports this insight.

To further validate our hypothesis of user convergence to the L2 feed, Table \ref{tab:ab_l1_l2_transition} shows a decrease in L1 depth (indicating the count of subsequent feed fetches) and an increase in L2 depth. Increment in L2 transition (\#times users switch from the L1 feed to the L2 feed) and decrease in S2L2 (Second to L2: time in seconds users take to open any post on L2 feed after session start) confirms higher and early user convergence to L2 feed. Including the feedback from the L2 feed improves user experience on the overall platform as indicated by the platform-level metrics in Table \ref{tab:ab_l1_l2_transition}, with metrics showing increased user retention, engagements, and interactions with the platform.

We also examined the composition of the suggested posts between the variants and observed that including the L2 feedback results in more suggestions from personalized candidate generators like field-aware factorization machines ~\cite{juan2016field} compared to non-personalized candidate generators (e.g. popularity). However, we did not observe any statistically significant difference in the genre distribution of the posts. Finally, we found that variant-2 performs better in most of the metrics compared to variant-1, which is consistent with the offline numbers we observed and our hypothesis of using the sum of feedback on the L2 feed being optimal compared to when discounting is used.

\begin{table}[!t]
\centering

\vspace{-2em}
\caption{Online \% gain in metrics when compared to the model trained only on L1 signal. All results are
statistically significant according to a 2-tailed t-test at $p<0.05$ after Bonferroni correction, except for those marked by$^+$.}

\begin{subtable}{\textwidth}
\centering
{
\caption{Engagement signals for L1 and L2 feed}
\label{tab:ab_feed_specific}

\begin{tabular}{ccccccc}

\toprule
Feed & variant & likes & shares & favs & dwell time & clicks \\
\midrule
\multirow{2}{*}{L1} & variant-1 & 5.9 & 4.3 & 5.7 & -1.0$^+$ & 5.1 \\
& variant-2 & 6.5 & 5.6 & 5.2 & -1.2$^+$ & 7.1 \\
\hline
\multirow{2}{*}{L2} & variant-1 & 3.6 & 1.1$^+$ & 3.1 & 1.5 & -\\
& variant-2 & 4.2 & 2.7 & 2.5 & 3.3 & - \\
\bottomrule
\end{tabular}
}
\end{subtable}

\begin{subtable}{\textwidth}
\centering
{
\caption{Platform \& transition metrics}
\label{tab:ab_l1_l2_transition}

\begin{tabular}{cccccccc}
\toprule
& \multicolumn{4}{c|}{L1 to L2 transition metrics} & \multicolumn{3}{c}{Platform level metrics} \\
\hline
variant & S2L2 & L1 depth & L2 depth & L2 transition & Engagements & \#Session & Retention  \\
\midrule
variant-1 & -0.9 & -1.43$^+$ & 2.53 & 8.15 & 0.43 & 0.3$^+$ & 0.12 \\
variant-2 & -2.1 & -4.14 & 5.42 & 11.23 & 0.71 & 0.5 & 0.17 \\
\bottomrule
\end{tabular}
}
\end{subtable}

\label{tab:online_ab}
\end{table}

%% file: 7_discussion.tex
\section{Discussion \& Future Work}
In this work, we have proposed a modeling framework for the nested feed structure that is prevalent in modern applications.
We provided a theoretical explanation for why optimizing for combined signals on both feeds is more effective than optimizing them independently. We considered different user feedback signals as the proper relevance signal. We used the DCG metric to demonstrate that incorporating L2 feedback in the L1 predictor reduces the loss in DCG compared to using the L1 predictor alone. Furthermore, we observed consistency between our online metric and the offline metric and observed improvement in both short-term engagement and user retention at the platform level. Our current work focused on optimizing the ranking policy for the L1 feed while keeping the L2 feed ranking policy fixed. We envision future extensions of this work where we can jointly optimize the L1, and L2 feeds using a single policy trained end-to-end.